\documentclass[12pt,preprint]{aastex}

\def\FLASH                 {{\sc flash}}
\def\PARAMESH              {{\sc paramesh}}
\newcommand{\orlando}{}
\newcommand{\orlandobis}{}

\shorttitle{Non-equilibrium of ionization in nanoflaring loops}
\shortauthors{Reale \& Orlando}

\begin{document}


\title{Non-equilibrium of Ionization and the Detection of Hot
       Plasma in Nanoflare-heated Coronal Loops}

\author{Fabio Reale\altaffilmark{1}}
\affil{Dipartimento di Scienze Fisiche \& Astronomiche, Universit\`a di
       Palermo, Sezione di Astronomia, Piazza del Parlamento 1, 90134 Palermo,
       Italy}
\author{Salvatore Orlando}
\affil{INAF - Osservatorio Astronomico di Palermo ``G.S.
       Vaiana'', Piazza del Parlamento 1, 90134 Palermo, Italy}


\altaffiltext{1}{INAF - Osservatorio Astronomico di Palermo ``G.S.
       Vaiana'', Piazza del Parlamento 1, 90134 Palermo, Italy}


\begin{abstract}
Impulsive nanoflares are expected to transiently heat the plasma confined
in coronal loops to temperatures of the order of 10 MK. Such hot plasma
is hardly detected in quiet and active regions, outside flares. During
rapid and short heat pulses in rarified loops the plasma can be highly out of
equilibrium of ionization. 
Here we investigate the effects of
the non-equilibrium of ionization (NEI) on the detection of hot
plasma in coronal loops. 
Time-dependent loop hydrodynamic simulations
are specifically devoted to this task, including saturated thermal
conduction, and coupled to the detailed solution of the equations of
ionization rate for several abundant elements.  In our simulations,
initially cool and rarified magnetic flux tubes are heated to 10 MK
by nanoflares deposited either at the footpoints or at the loop apex.
We test for different pulse durations, and find that, due to NEI effects,
the loop plasma may never be detected at temperatures above $\sim 5$ MK for
heat pulses shorter than about 1 min.
We discuss some implications in the framework of
multi-stranded nanoflare-heated coronal loops.
\end{abstract}




\section{Introduction}
\label{sec1}

Nanoflares -- small scale highly transient heating episodes --
are among the main candidates as source of coronal loop heating
(e.g., \citealt{1988ApJ...330..474P}; \citealt{1994ApJ...422..381C};
\citealt{2006SoPh..234...41K} and references therein). The conjecture is still under debate
because nanoflares have been hardly detected so far. There are
many possible reasons for this difficult detection. For instance,
a very frequent occurrence may inhibit the resolution of the single
event. Also the efficient thermal conduction and low emission measure
during the pulses can reduce and delay the signatures of the heating
(e.g., \citealt{1987ApJ...312..895P}; \citealt{1995A&A...299..225R}). Also very small
pulses distributed in a very finely structured loop may be difficult
to detect. One of the main arguments invoked as crucial evidence of
nanoflare heating is the detection of high temperature ($10$ MK)
plasma components in observations of coronal loops (\citealt{1997ApJ...478..799C}; \citealt{2006SoPh..234...41K}), outside of
proper flares. The presence of such hot component is often predicted
by hydrodynamic modeling of coronal loops heated by transient pulses
(\citealt{2005ApJ...628.1023P}; \citealt{2006ApJ...647.1452P}; \citealt{2004ApJ...605..911C}). In particular any
heat spike able to bring the loop to the observed brightness should
be so intense as to heat the plasma to temperatures of the order of
$10$ MK at least for a transient time interval. In the hypothesis of a
finely structured loop where a whole distribution of heat pulses occur
continuously, such plasma might be detectable, at least as a hot tail in
the emission measure distribution. The evidence of hot plasma has been
difficult so far. For instance, a dominant plasma component at about 3 MK
is shown by recent thermal maps of active regions obtained from wide-band
multi-filter imaging observations (\citealt{2007Sci...318.1582R}) with
the X-Ray Telescope (XRT, \citealt{2007SoPh..243...63G}) on board the
Hinode mission (\citealt{2007SoPh..243....3K}).

There are possible explanations also for the difficult detection of hot plasma. The most immediate one
is linked to the inertia of the plasma dynamics.  A heat pulse deposited
in a coronal loop drives evaporation of chromospheric plasma.  The loop
is filled with hot and dense plasma which makes it bright in the X-rays.
For short heat pulses, the plasma may evaporate from the
chromosphere on time scales longer than the heat duration. Therefore
the heated strand may only later acquire enough emission measure to become
visible, while it is cooling. However, in short loops ($\sim 10^9$
cm) the sound crossing time is of the order of one minute (e.g.,
\citealt{2007A&A...471..271R}) and we may expect to detect hot spots
within less than half a minute.

There is another less obvious effect which may make the detection of hot
plasma harder: the time lag of the plasma to change its ionization
from a cool to a hot state. 
An impulsive energy input drives plasma thermal and dynamic changes on
relatively
short timescales. Electron
excitation, de-excitation, ionization and recombination processes of the 
ion species have
other timescales. If the timescale, for instance,
of the temperature evolution is much shorter than the ionization and
recombination timescales, the degree of ionization can be very
different from the equilibrium conditions corresponding to the local
electron temperature (\citealt{1977ApJ...217..621S}; \citealt{1999A&A...346.1003O}; \citealt{2006SoPh..234...41K}). So, during a fast
temperature increase, the plasma ions can be at a lower ionization state
than the equilibrium state corresponding to the instantaneous temperature.
Such non-equilibrium of ionization (NEI) effects may become important in
the interpretation of what we observe.  If the heat pulse is sufficiently
short, the ions and their emitted radiation may not even have enough
time to ``sense" the hot temperature status, they would adjust to the
temperature variations deep in the cooling phase, and we would detect
radiation from cooler ion conditions at any time.

NEI effects in nanoflare-heated loops have already been investigated in
the past. \cite{1982ApJ...255..783M} found that NEI can significantly alter
the relative ionic abundances in the quiet corona. \cite{1989SoPh..122..245G}
examine the effect of NEI on the observability of coronal variations.
\cite{2003A&A...407.1127B}
found that NEI can modify considerably the radiative losses
function. \cite{2003A&A...411..605M} investigated the role of NEI in the
transition region brightenings driven by loop condensations.
\cite{2006A&A...458..987B} model nanoflare heating in coronal loops 
and remark the importance of NEI effects, the presence of hot plasma with
low emission measure, and Doppler-shifts as possible diagnostics of the heating.

Here we investigate the effect of NEI during nanoflaring activity on
the detectability of hot plasma in coronal loops. We will
model a coronal loop strand heated by nanoflares of different durations,
compute the corresponding evolution of the relevant ion species, and
compare it with the evolution in full ionization equilibrium.
We will evaluate the effects on the expected temperature distribution of the
loop emission measure, which will tell us about the existence of 
significantly emitting hot plasma components.
This will allow us to put constraints on the nanoflare characteristics
which lead to the observability of hot plasma and indirectly also
on the fine loop structuring.
In Sec.\ref{sec:model} the modeling is described, in Sec.\ref{sec:result} the results are illustrated and discussed in Sec.\ref{sec:discus}.

\section{Modeling}
\label{sec:model}


Our model is set up to explore the conditions for plasma confined in a
coronal magnetic flux tube and heated to $10$ MK by nanoflares to be
detectable as hot plasma considering the effects of NEI.
The coronal loop contains low $\beta$ plasma, and, as customary for
standard loop models, we assume that plasma moves and transports energy
only along the magnetic field lines, so that a one-dimensional description
is adequate (e.g., \citealt{1982ApJ...252..791P}). The model takes into
account the gravity stratification, the thermal conduction (including
the effects of heat flux saturation), the radiative losses, an external
heating input, and the NEI effects. The following fluid equations
of mass, momentum, and energy conservation are solved, considering
only the relevant components along the loop magnetic field lines:

\begin{equation}
\frac{\partial \rho}{\partial t} + \nabla \cdot \rho \mbox{\bf v} = 0
\label{eq1}
\end{equation}

\begin{equation}
\frac{\partial \rho \mbox{\bf v}}{\partial t} +\nabla \cdot \rho
\mbox{\bf vv} + \nabla P = \rho\mbox{\bf g}
\end{equation}

\begin{eqnarray}
\lefteqn{\frac{\partial \rho E}{\partial t} +\nabla\cdot (\rho
E+P)\mbox{\bf v} =} \nonumber \\
 &  \rho \mbox{\bf v}\cdot \mbox{\bf g} -\nabla\cdot q
+ Q(s, t)-n_e n_H \Lambda(T)
\end{eqnarray}

\[
\mbox{where \hspace{0.5cm}} E = \epsilon +\frac{1}{2} |\mbox{\bf
v}|^2~,
\]

\noindent
is the total gas energy (internal energy, $\epsilon$, and kinetic
energy), $t$ is the time, 
$s$ is the coordinate along the loop,
$\rho = \mu m_H n_{\rm H}$ is the mass density,
$\mu = 1.26$ is the mean atomic mass (assuming solar abundances),
$m_H$ is the mass of the hydrogen atom, $n_{\rm H}$ is the hydrogen
number density, $n_{\rm e}$ is the electron number density, {\bf v}
the plasma flow speed, $P$ the pressure, {\bf g} the gravity,
$T$ the temperature, $q$ the conductive flux,
$Q(s,t)$ a function describing the transient input heating, $\Lambda(T)$
is the radiative losses per unit emission measure (e.g. \citealt{rs77};
\citealt{mgv85}; \citealt{2000adnx.conf..161K}). Here we consider the
radiative losses function in equilibrium of ionization. Although this 
assumption makes our
description not entirely self-consistent, it does not affect 
significantly our results, 
because in the critical evolution phases the energy losses are dominated
by transport by thermal conduction (see Sec.\ref{sec:sim}). We use the ideal gas
law, $P=(\gamma-1) \rho \epsilon$, where $\gamma=5/3$ is the ratio of
specific heats.

\orlando{The set of the continuity equations for each ion species is
expressed as:}

\begin{equation}
\frac{\partial n_i^Z}{\partial t} + \nabla \cdot n_i^Z \mbox{\bf v} =
R_i^Z ~~~~~~~~\begin{array}{l}(Z = 1, ..., N_{elem})\\\\
(i = 1, ..., N_{ion}^Z) \end{array}
\label{eq4}
\end{equation}

\noindent
\[
\mbox{where}~~~R_i^Z = n_e [n_{i+1}^Z\alpha_{i+1}^Z + n_{i-1}^Z
S_{i-1}^Z - n_i^Z(\alpha_i^Z+S_i^Z)]~,
\]

\noindent
\orlando{$n_{i}^Z$ is the number density of the $i$-th ion of the element $Z$,
$N_{elem}$ is the number of elements, $N_{ion}^Z$ the number of ionization
states of element $Z$, $\alpha_i^Z$ are the collisional and dielectronic
recombination coefficients, and $S_i^Z$ the collisional ionization
coefficients (\citealt{summers74}).}

Given the importance of rapid transients, fast dynamics and steep
thermal gradients in this work, we consider both the classical and saturated
conduction regimes.
To allow for a smooth transition between them, 
we follow \citet{1993ApJ...404..625D} and define
the conductive flux as

\begin{equation}
q = \left(\frac{1}{q_{\rm spi}}+\frac{1}{q_{\rm sat}}\right)^{-1}~.
\end{equation}

\noindent
Here $q_{\rm spi}$ represents the classical conductive flux
(\citealt{spi62})

\begin{equation}
q_{\rm spi} = -\kappa(T)\nabla T
\label{spit_eq}
\end{equation}

\noindent
where the thermal conductivity is $\kappa(T) = 9.2\times 10^{-7}
T^{5/2}$
erg s$^{-1}$ K$^{-1}$ cm$^{-1}$. The saturated flux,
$q_{\rm sat}$, is (\citealt{cm77})

\begin{equation}
q_{\rm sat} = -\mbox{sign}\left(\nabla T\right)~ 5\phi \rho c_{\rm
s}^3,
\label{therm}
\end{equation}

\noindent
where $c_{\rm s}$ is the isothermal sound speed, and $\phi$ is a correction
factor
of the order of unity. We set $\phi = 1$ according to the values
suggested for the coronal plasma (\citealt{1984ApJ...277..605G};
\citealt{1989ApJ...336..979B}, \citealt{2002A&A...392..735F}, and
references therein).

The transient input heating is described empirically as a separate
function of space 
and time:

\begin{equation}
Q(s,t) = H_0 \times g(s) \times f(t)
\end{equation}

\noindent
where $g(s)$ is a Gaussian function:

\begin{equation}
g(s) = \exp[-(s-s_0)^2/2\sigma^2]~,
\end{equation}

\noindent
and $f(t)$ is a pulse function:

\begin{equation}
f(t) = \left\{\begin{array}{ll}
0, & t \leq  0 \\ \\
1, & 0 <t\leq t_H \\ \\

0, & t > t_H
\end{array}\right.
\end{equation}

The calculations described in this paper were performed using the 1-D version
of the \FLASH\
code (\citealt{for00}), an adaptive mesh refinement multiphysics code. For
the present application, the code has been extended by additional
computational modules to handle the plasma thermal conduction
(see \citealt{2005A&A...444..505O} for details of the
implementation), the NEI effects, the radiative losses, and the
heating function. The implementation of the description of the ionization
balance in the \FLASH\ code is described in the Appendix.


We consider an initially cool and rarified semicircular
magnetic flux tube (loop or loop strand)
with half-length $L = 3 \times 10^9$ cm, and uniform cross-section
area. The area appears only as a multiplicative factor in this description
and the size is typical of active region loops. The loop is initially
at equilibrium according to loop scaling laws (\citealt{1978ApJ...220..643R})
and to hydrostatic conditions 
(\citealt{1981ApJ...243..288S})
assuming it lies on a plane vertical to the solar
surface. 
The initial base pressure is $p_0 = 0.055$
dyn/cm$^2$, corresponding to a maximum temperature $T_0 = 0.79$ MK, at
the loop apex. The corona is linked with a steep transition region to
an isothermal chromosphere uniformly at $T_c = 20000$ K and $0.5 \times
10^9$ cm thick. We assume that the loop is symmetric with respect
to the vertical axis across the apex, and therefore we simulate only half of 
the loop, with a total extension of 
$3.5 \times 10^9$ cm.
\orlando{All the ion species are assumed initially in
equilibrium of ionization in the whole computational domain.}

\orlando{At the coarsest resolution, the adaptive mesh algorithm used
in the \FLASH\ code (\PARAMESH; \citealt{mom00}) uniformly covers the
computational domain with a mesh of $16$ blocks, each with $8$ cells.
We allow for 3 levels of refinement, with resolution increasing
twice at each refinement level. The refinement criterion adopted
(\citealt{loehner}) follows the changes in density and temperature. This
grid configuration yields an effective resolution of $\approx 3.4\times
10^6$ cm at the finest level, corresponding to an equivalent uniform
mesh of $1024$ grid points. We use fixed boundary conditions at $s=0$
and reflecting boundary conditions at $s=s_{\rm max}$ (consistent with
the adopted symmetry).}

\section{Results}
\label{sec:result}

\subsection{The Simulations}
\label{sec:sim}

In the loop outlined above we inject one nanoflare as intense as to heat the plasma
and keep it at $\approx 10$ MK, if the heating were steady. The choice of
the parameters is dictated by our scope of exploring the influence
of NEI on the detectable thermal conditions. In this perspective,
the pulse duration becomes the critical parameter: if the heat pulse
is short enough, the plasma may not have enough time to adjust to 
ionization conditions appropriate of 10 MK, before the heating is
off. Therefore, we consider longer and longer heat pulse durations with a
logarithmic sampling, i.e. $t_H = 5$ s, 30 s, 180 s.  For each of these
durations we consider two possible locations of the pulse depositions:
at the footpoints, namely $s_0 = 8 \times 10^8$ cm from the base of the
chromosphere (i.e. $3 \times 10^8$ cm from the base of the corona) and at
the apex, namely $s_0 = 3.5 \times 10^9$ cm.  The spatial width of the
pulse is smaller at the footpoints ($\sigma = 10^8$ cm) and larger at
the apex (\orlando{$\sigma = 5\times 10^8$ cm}). To have the same total energy
deposition rate, the maximum volume deposition
rates are \orlando{$H_0 = 1.5$} erg cm$^{-3}$ s$^{-1}$ and \orlando{$H_0 =
0.2$} erg cm$^{-3}$ s$^{-1}$, respectively.

We come out with a total of six simulations. They are carried out for
a time which covers the pulse duration and a couple of loop cooling
times:

\begin{equation}
\tau_{dec} = 120 \frac{L_9}{\sqrt{T_7}} \approx 360 ~~ \rm s
\end{equation}

\noindent
where $L_9 = 3$ is $L$ in units of $10^9$ cm and $T_7 \approx 1$ is the
loop maximum temperature in units of $10$ MK (\citealt{1991A&A...241..197S}). 
All simulations then span a time interval of about \orlando{800} s.


The evolution of nanoflaring plasma confined in coronal loops is
well-known from previous work (\citealt{1993pssc.symp..151P};\citealt{2002ApJ...579L..41W};\citealt{2003ApJ...593.1174W};\citealt{2005ApJ...628.1023P};\citealt{2005ApJ...622..695T})
and is on a smaller scale similar to that
of properly flaring loops (e.g. \citealt{1980SoPh...68..351N};\citealt{1982ApJ...252..791P};\citealt{1983ApJ...265.1103D};\citealt{1984ApJ...279..896N};\citealt{1985ApJ...289..414F};\citealt{1987ApJ...312..895P};\citealt{1995A&A...299..225R}). 
As additional feature, our simulations include the effect of saturated
thermal conduction, which might be important here since we study the
evolution on small time scales (see also \citealt{2003SPD....34.1006K}).
In the following, we just draw a basic outline of the results adapted to our simulations. 

\begin{figure*}[!t]
  \centering \includegraphics[width=15cm]{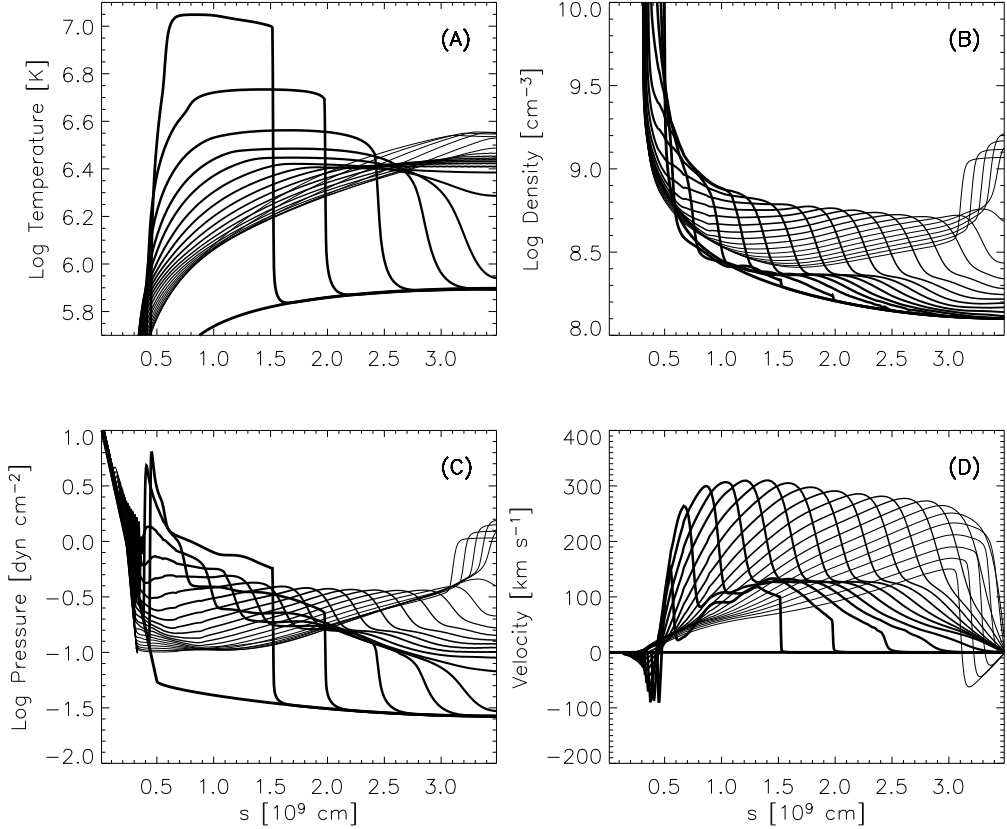}
  \caption{Evolution of the plasma temperature (A), density (B), pressure
  (C), and velocity (D) distributions along the loop from the chromosphere
  to the coronal loop apex, sampled every 5 s from 0 to
  100 s (thinner and thinner lines), for heat pulse location at the loop base and heat pulse
  durations $t_H = 5$ s.}
  \label{fig0}
\end{figure*}

Let's
consider the simulation with $t_H = 5$ s and heat pulses deposited at the
footpoints (Fig.~\ref{fig0}). The sudden heat deposition determines a
local increase of temperature (to $\sim 10$ MK) and pressure (to $\sim
1$ dyn/cm$^2$).  A fast thermal front propagates upwards along the
loop. The cool chromosphere is heated and expands upwards with a strong
evaporation front. By the time the thermal front has reached
$s\sim 1.5 \times 10^9$ cm and the evaporation front $s\sim 6\times10^8$
cm, the heat pulse is already over. Due to efficient thermal conduction,
the plasma then immediately begins to cool, already during the propagation of
the thermal front, and the maximum temperature rapidly decreases to $\sim
2.5$ MK in less than a minute.  The thermal front reaches the apex, and
therefore the loop thermalizes, in $\sim 30$ s. The impulsive evaporation
front moves at velocity of about 300 km/s (Fig.\ref{fig1}D)
and fills the loop in $\sim 75$
s. The density keeps on increasing throughout the loop with more gentle
fronts (the peak velocity rapidly decreases to less than 200 km/s) for
further tens of seconds. Meanwhile the plasma accumulates at the apex
reaching a density of about $10^{9}$ cm$^{-3}$. The compression heats
the plasma again above 3 MK at the apex.  Then the plasma begins to drain
and the density to decrease (not shown), following the radiation cooling
time (\citealt{2004ApJ...605..911C}; \citealt{2007A&A...471..271R}).

The overall evolution does not change much in the other
simulations. Longer heat pulses drive more and longer plasma
evaporation. Pulses deposited at the apex produce downward thermal fronts
but as soon as the thermal front hits the chromosphere, it drives an
evaporation front similar to that driven by the pulses at the footpoints.

\subsection{Non-equilibrium of Ionization}

\begin{figure}[!t]
  \centering \includegraphics[width=8.5cm]{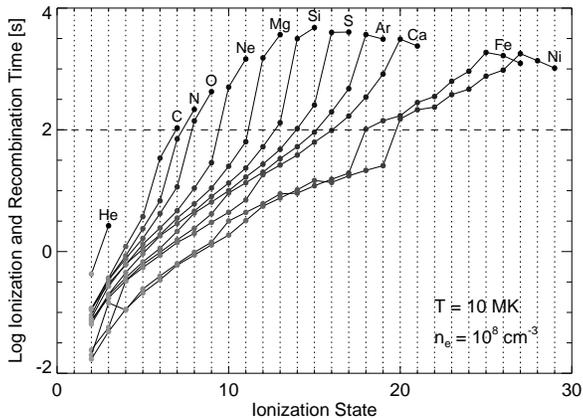}
  \caption{Ionization/recombination timescale (in log scale)
   for various ion species (dots) vs the ionization state for the labeled
   elements, computed for a temperature $10$ MK and a density $10^8$
   cm$^{-3}$. The dashed horizontal line marks 100 s.}
  \label{fig2}
\end{figure}

The importance of highly transient processes such as NEI is basically
dictated by their timescales related to the timescales of the dynamics
and heating/cooling driven by the nanoflares. Fig.~\ref{fig2} shows the
combined ionization/recombination timescale for the {\it i}-th ion species
of the element $Z$, derived from Eq.(\ref{eq4}) as

\begin{equation}
\tau_{\rm i}^Z = \frac{1}{n_{\rm e}(\alpha_{\rm i}^Z+S_{\rm i}^Z)}~,
\end{equation}

\noindent
for twelve important
elements, computed for a temperature $10$ MK and a density $10^8$
cm$^{-3}$. \orlandobis{Analogous timescales have been provided, in
tabular form, by \citet{1989SoPh..122..245G} in the study of observable
variability of spectral lines in soft X-ray and XUV regions of the
solar corona.}

Clearly, $\tau_{\rm i}^Z$ increases with the ion species
and becomes larger than 100 s for the highest ion species of C, N, O,
Ne, Mg, Si, S, Ar and Ca. More species are involved in high ionization
times for Fe and Ni. The modeling shows that the plasma
immediately cools down considerably as soon as the heat pulse stops.
As a consequence, from 
Fig.~\ref{fig2} we expect that the high ion species will
have no time to adjust to high temperature status if the heat pulse
lasts significantly less than 100 s.

\orlandobis{The ionization and recombination timescale is dominated by
either ionization or recombination rate, depending on the sign of the 
temperature jump
and on the ionization state of the element considered: in general,
$\tau_{\rm i}^Z$ is dominated by the ionization rate for the lowest ion
species and by the recombination rate for the highest ion species. The
ionization state for which ionization and recombination rates are
comparable depends on the temperature: the higher the temperature,
the higher is this ionization state. In the case considered here,
$\tau_{\rm i}^Z$ is dominated by the ionization rate for most of the ion
populations (see also \citealt{2006ApJ...647.1452P}) and by recombination
rate for the top ionization states (e.g. Si XV, S XVI, Ca 20,
Fe XXVI). As shown later, we found the effective ionization
temperature generally lower than the electron temperature, even when
the plasma is cooling; as a consequence, in the case of nanoflares
discussed here, $\tau_{\rm i}^Z$ is generally dominated by the ionization
rate for most of the ion populations.}

\begin{figure*}[!t]
  \centering \includegraphics[width=16cm]{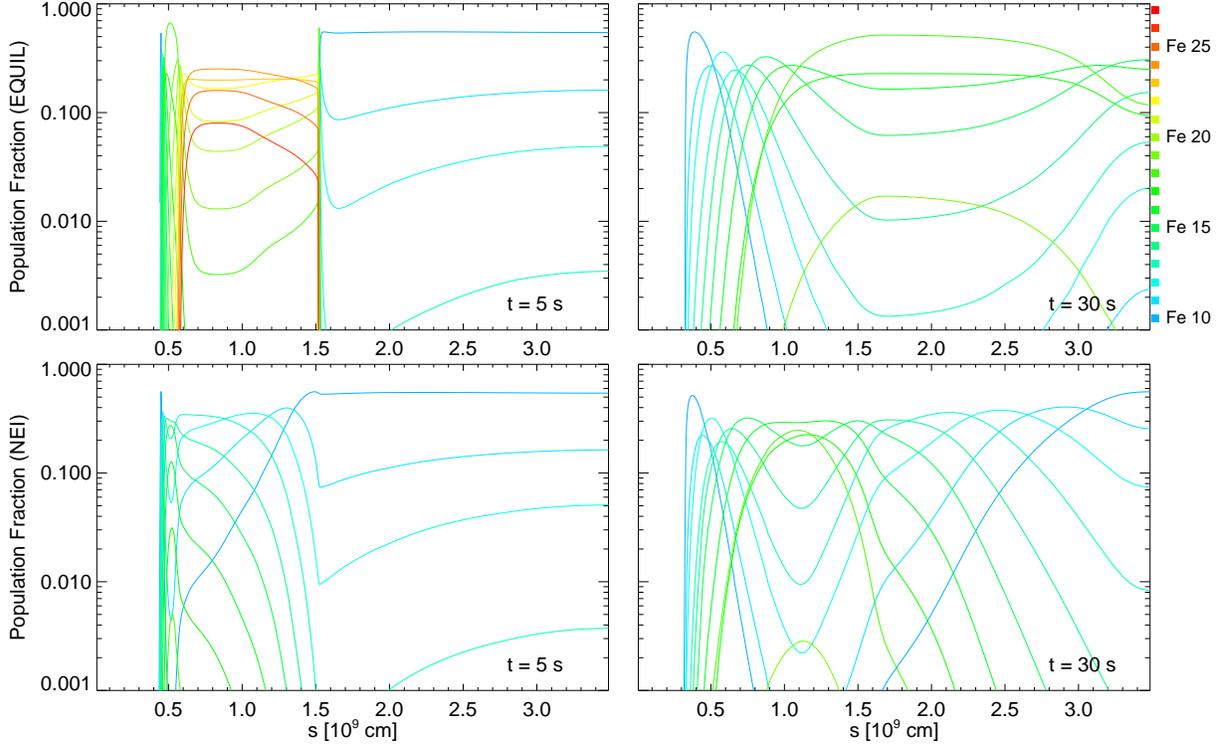}
  \caption{Distributions of Fe population fractions (color-coded on the
right axis of the upper right panel) along the
  loop from the chromosphere to the coronal loop apex, assuming
  equilibrium ionization (upper panels) or considering the deviations
  from equilibrium ionization (lower panels), at the labeled
  times (5 s and 30 s), for the simulation with heat pulses located at the loop base and heat pulse
  duration $t_H = 5$ s.}
  \label{fig1}
\end{figure*}

\orlando{As an example, Fig.~\ref{fig1} shows the distribution of
population fractions of Fe along the loop derived assuming equilibrium
ionization (upper panels) and considering the deviations from equilibrium
ionization (lower panels) for the simulation 
with heat pulses deposited at the footpoints
lasting $t_H = 5$ s
(see also Fig.~\ref{fig0}). During
the early phase of the evolution, the deviations from equilibrium of
ionization are very large at the loop footpoints due to the sudden 
local increase
of temperature to $\sim 10$ MK (see left panels of Fig.~\ref{fig1}). 
Whereas significant populations are expected above Fe XX low in the 
loop (below $1.5 \times 10^9$ cm, reddish lines in the upper left panel),
no such highly ionized Fe species appears when NEI is taken into account.
As
expected, the high ion species are not able to adjust to the plasma
temperature before the heat pulse is over.  
Then the plasma cools
down reducing the deviations from equilibrium ionization. 
The deviations are still significant later at $t = 30$ s, when the loop
has thermalized (see Fig.~\ref{fig0}) and the maximum temperature is $\sim
3$ MK. At this time, no highly ionized Fe species is present both
considering and not considering NEI, because of the effective cooling.
}

\begin{figure*}[!t]
  \centering \includegraphics[width=13cm]{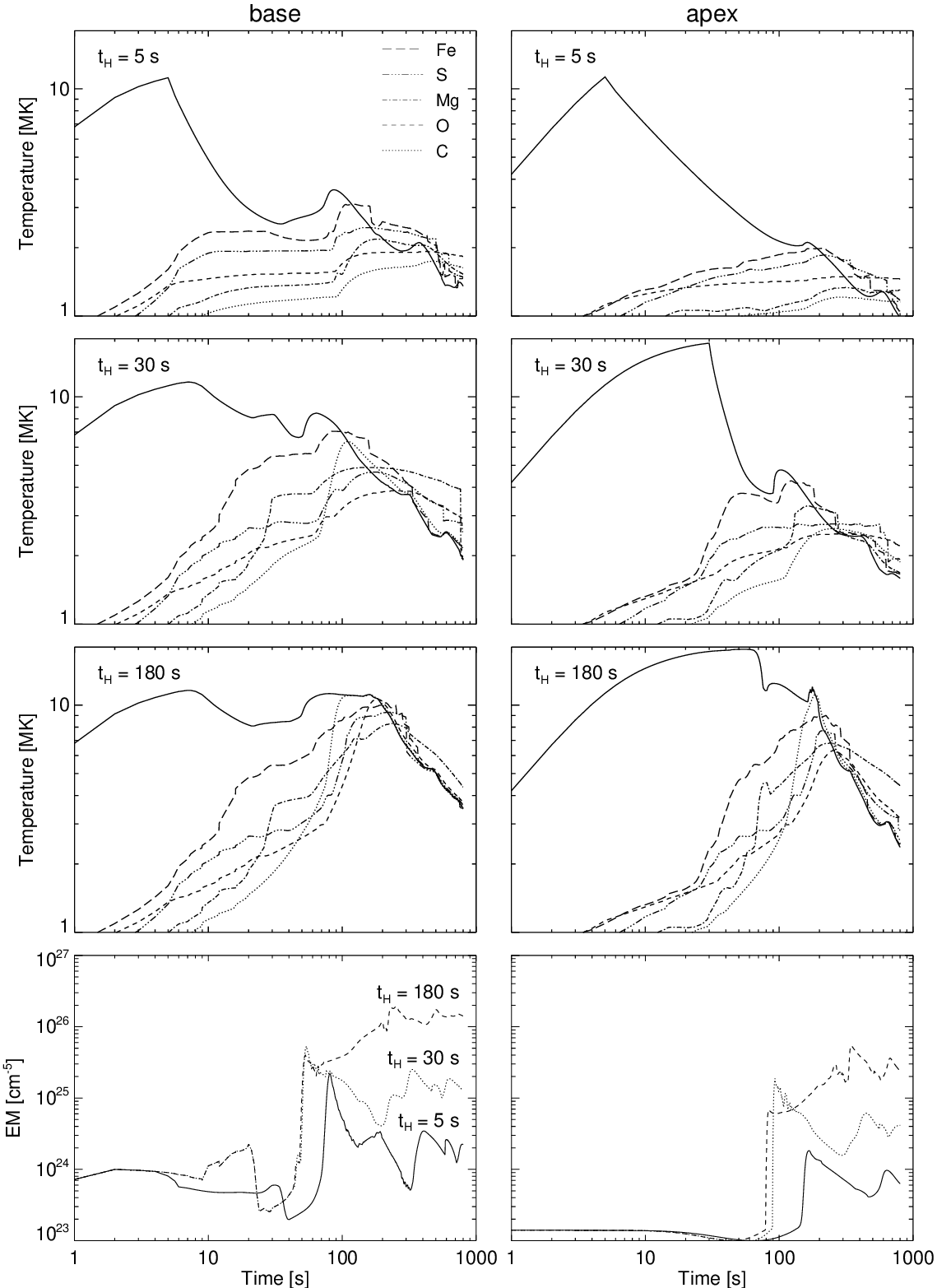}
  \caption{\orlando{Evolution of the maximum electron temperature (solid
   lines) and of the maximum ``NEI" temperature (see text for definition) for
the labelled elements (dashed and dotted lines). 
The figure shows the results for
   heat pulse location at the loop base (left panels) and at the apex (right
panels)
   and for heat pulse durations $t_H = 5$ s, 30 s, and 180 s.
Bottom panels: evolution of the coronal emission
   measure at the maximum plasma temperature. The time is in log scale to 
zoom up the fast initial evolution.
}}
  \label{fig3}
\end{figure*}

This time lag of the evolution of the ion species has important
implications on the temperature diagnostics of the plasma from its
emission. To address this issue, from the modeled population fractions,
we derive the temperature which best matches the actual ionization state 
of the ion species for each of the elements in our simulations: 
\orlandobis{operatively this is found as the one of the equilibrium temperatures with the 
most similar three most populated ionized states.
It is worth noting that
at least two population fractions are necessary to find a
unique value of temperature and that this ``NEI" temperature should not
be intended as an exact temperature but as the temperature which best describes
the ionization state.} This is done at each time and position along
the loop.  Fig.~\ref{fig3} shows the evolution (on a logarithmic time
scale) of the plasma maximum temperature and of the maximum NEI temperature
for five representative elements (C, O, Mg,
S, Fe). In the same figure the evolution of the coronal emission measure
at the maximum plasma temperature is also shown for comparison.  The ion
species adjust to a hotter status very gradually, on a time scale of about
100 s (Fig.~\ref{fig2}).  Since this time is slightly longer than the
time taken by the emission measure to increase significantly (as shown
by the bottom panels), we expect observable effects in the X-ray band.
For $t_H = 5$ s, despite the maximum electron temperature is above 10 MK,
all the ion species are never represented by a temperature larger than
about 3 MK all over the event, for any pulse location. For $t_H = 30$
s, Fe (and Ni) reaches a temperature about $7-8$ MK for pulses located
at the footpoints and about 6 MK for pulse deposited at the apex. This
occurs for a very limited time range (between 100 and 200 s). The other
elements mostly stay at temperatures well below 5 MK. The results change
quite substantially for $t_H = 180$ s. Although with delay ($\ge 100$
s), here many elements have enough time to adjust to ionization stages
typical of higher temperatures, between 7 MK and 10 MK, for both heat
pulse locations. The high ionization state is maintained for longer time
intervals (more than 200 s).

\begin{figure*}[!t]
  \centering \includegraphics[width=14cm]{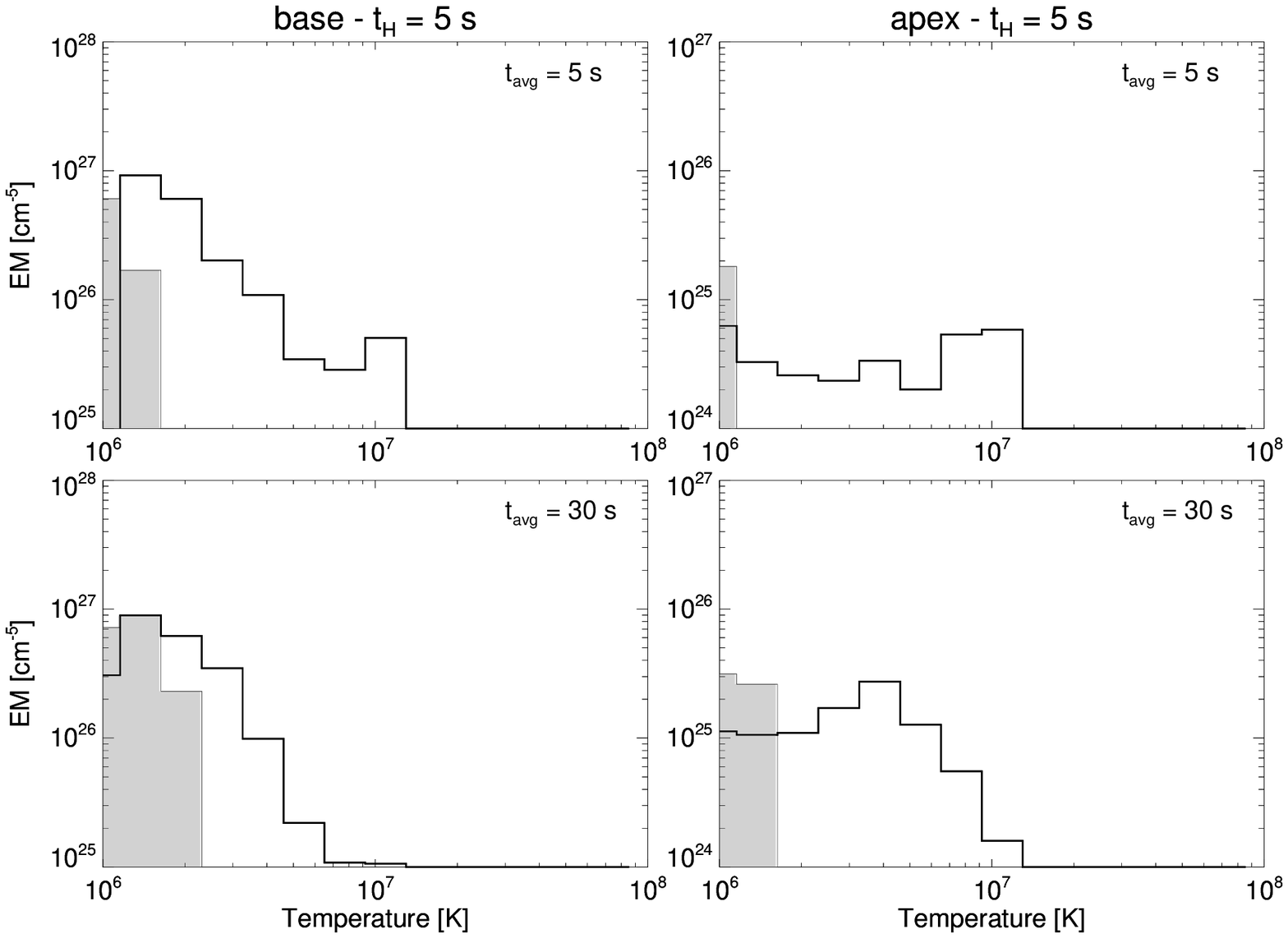}
  \caption{\orlando{Distributions of emission measure vs temperature
  averaged over different time intervals obtained either taking (thick
  lines) or not taking (shaded areas) the deviations from equilibrium
  ionization into account (see text) for heat pulse location at the loop
  base (left) or at the apex (right) and heat pulse duration $t_H = 5$
  s.}}
\label{fig4}
\end{figure*}

\begin{figure*}[!t]
  \centering \includegraphics[width=14cm]{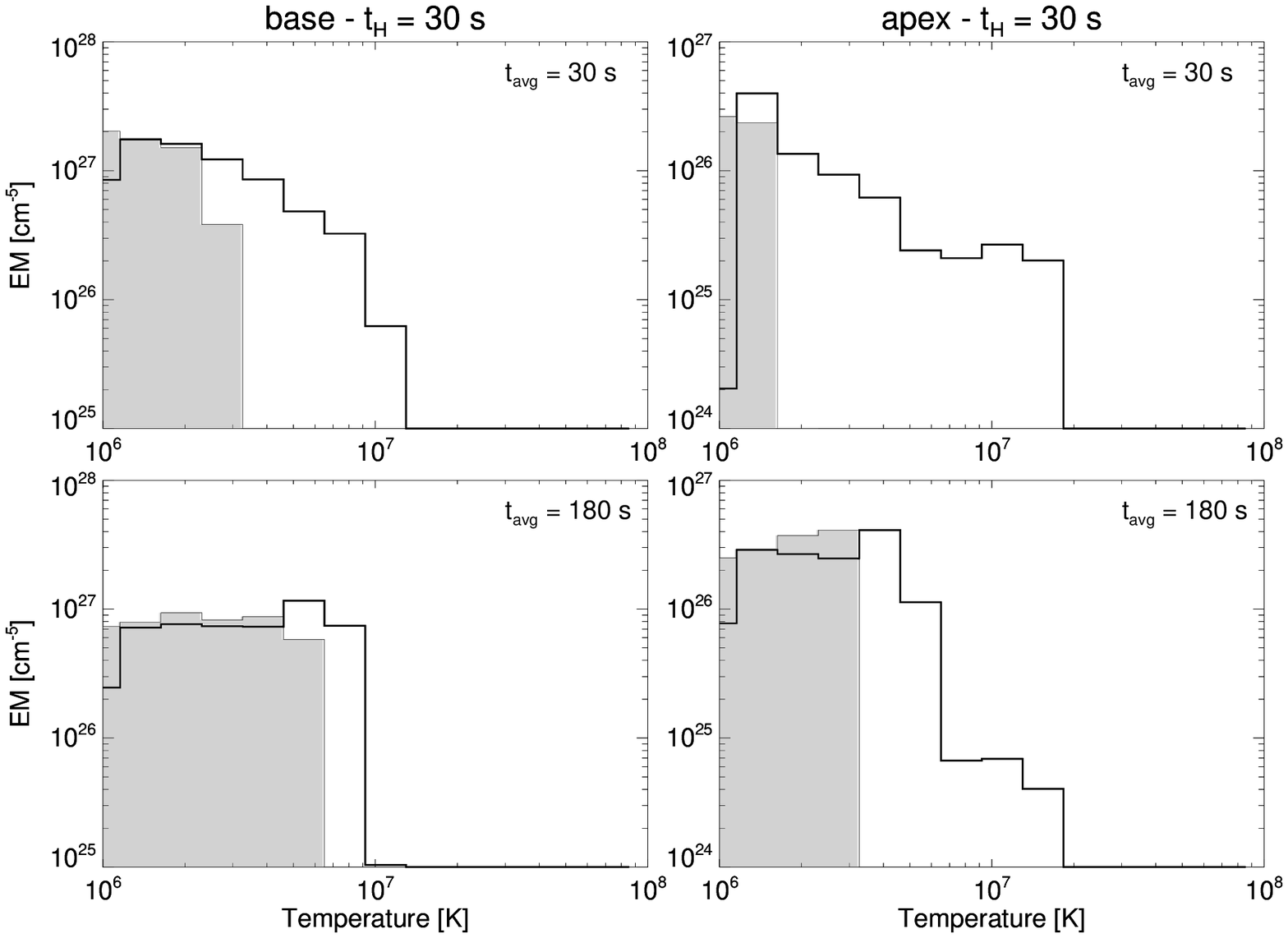}
  \caption{\orlando{Same as Fig. \ref{fig4}, for heat pulse duration $t_H =
  30$ s.}}
  \label{fig5}
\end{figure*}

\begin{figure*}[!t]
  \centering \includegraphics[width=14cm]{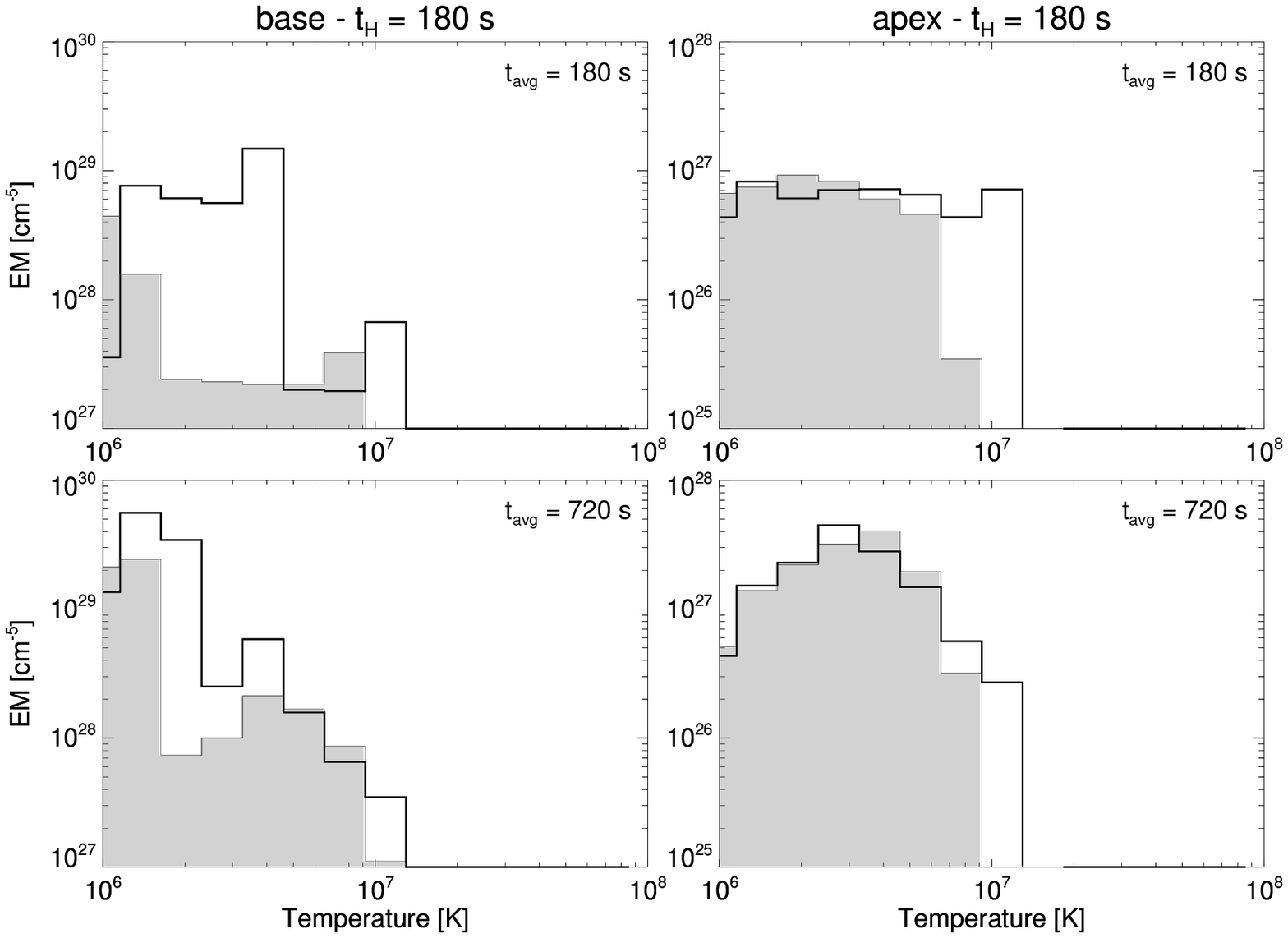}
  \caption{\orlando{Same as Fig. \ref{fig4}, for heat pulse duration $t_H =
  180$ s.}}
  \label{fig6}
\end{figure*}

The effect of long-lasting NEI is illustrated synthetically in Figs.
\ref{fig4}-\ref{fig6} which shows the total loop emission measure
distribution with temperature, EM($T$), averaged over different time
intervals. \orlando{The figure compares EM($T$) obtained either taking
or not taking the deviations from equilibrium ionization into account;
the former is obtained using the average ``NEI" temperature among those of
all the elements considered, the latter using directly the electron
temperature.  }

Averaging over the heat pulse duration only, we can see that EM($T$)
with and without NEI are invariably very different. The shorter the
pulse duration, the more the hot components are missing from the NEI
EM($T$) distributions. For $t_H = 5$ the hottest components including
NEI are only at about 1 MK, which increases to 3 MK for $t_H = 30$
s. The difference is less considerable for $t_H = 180$ s, but even in
this case the hot components are hardly above 5 MK, against 10 MK
without NEI. The differences are unsurprisingly reduced when we average over
longer times, which should approach a situation closer to the realistic
observations. Over a few durations of the heat pulses the differences are
still significant for the shortest heat pulses: for $t_H = 5$ s,
with NEI we do not find
components above 2 MK, whereas without NEI there are components almost to
10 MK. For $t_H = 30$ s the two EM($T$) distributions are more similar,
but with NEI we do not find components hotter than 6 MK. For the longest
pulse duration, the EM(T) distributions almost coincide.

\section{Discussion and conclusions}
\label{sec:discus}

The target of this work is to explore the effect of non-equilibrium
of ionization (NEI) in
nanoflare-heated loops.  
The importance of NEI effects on coronal observations had been already
pointed out and evaluated by \cite{1989SoPh..122..245G}. 
That study was devoted more specifically
to the diagnostics of rapid variability. Here, we focus on the effects on
the detection of hot plasma in nano-flaring loops.
To this
purpose we set up hydrodynamic simulations of plasma confined in
an active region loop heated to $\sim 10$ MK by more or less long
nanoflares. As new achievements specifically set up for this study,
we have included the effect of the saturated thermal conduction in the
modeling (\citealt{2003SPD....34.1006K}), and computed the evolution of the ion population fractions
of several important elements driven by a short heat pulse. Each
ion species takes a characteristic time to reach equilibrium conditions
for a certain plasma temperature and density. If the plasma conditions
vary on very small time scales, the ions are unable to adjust to
the new and rapidly changing conditions and the emitted spectrum during the
plasma variations can become very different from the one expected in
equilibrium conditions. If the variation is a sudden heating followed by
a sudden cooling, the emission may never or only partially adapt to the
transient hot conditions. To evaluate the importance of this effect and
its dependence on the heat pulse parameters, we have computed the
actual temperature of the plasma as it would appear in the radiation spectrum
according to the effective ionization state.

We find that the effects are significant if the heat pulses last less than
a minute or so. For durations of a few seconds the spectra will never show
temperatures higher than 2-3 MK even though the electron temperature
overcomes 10 MK. This effect is little dependent on the location of
the heat pulse, except for details. 
These results can be considered valid in wide
generality in spite of some limitations of our approach. Our assumption
of pulses leading to 10 MK is, of course, arbitrary, although this has
been taken as possible typical temperature in other works (\citealt{2006ApJ...647.1452P}). Conclusions are certainly valid for less intense pulses,
which heat less the plasma. We do not expect dramatic differences also for
more intense pulses (e.g. 20 MK), because the thermal conduction becomes
even more efficient and the plasma initial cooling (after the end of the
pulse) will be faster. Also the details of the shape of the heat pulse
should not have much influence, provided that the overall duration is
the same. In general, of course, more gentle heating should make plasma
reach equilibrium conditions more easily, while the opposite occurs for
more spiky pulses. More important is instead our assumption of a single
pulse inside a given thread. With it, we are excluding a repetition of
the pulse inside the same thread, or, at least, a very long repetition
time, so long as to have a rarified loop ($\sim 10^8$ cm$^{-3}$) again
before the new ignition. More frequent pulses would be released in a denser
plasma which is faster to adjust to ionization equilibrium, and in which,
therefore, we expect less NEI effects. The ionization times in fact scale
inversely with plasma density. Heat pulses released in a $\sim 10^9$
cm$^{-3}$ dense plasma, i.e. about ten times denser than our initially
rarified flux tube, would excite highly ionized element species about
ten times more rapidly, with typical times of about ten seconds
(Fig.~\ref{fig2}).

In the range $1-10$ MK the plasma emits radiation mostly in the X-ray
band and the spectra are dominated by the emission lines produced by
the recombination and deexcitation of highly ionized atoms (e.g.,
\citealt{1971ApJ...168..283T}, \citealt{1976ApJS...30..397K}). For
instance, hundreds of lines of at least ten Fe ion species populate
the spectra of plasma in that range of temperature. For this reason,
the ionization status of the emitting ions becomes extremely important
in the thermal ``appearance'' of the plasma, which we detect through
its radiation with remote sensing, while the continuum electron
bremsstrahlung emission -- which adjusts immediately to thermal and
dynamic plasma variations -- becomes more important at much higher
temperatures. Our results therefore apply both to the analysis of single
lines through high resolution spectroscopy, and to the diagnostics
from wideband multi-filter instruments in the X-ray and EUV bands,
e.g. Hinode/XRT or the Atmospheric Imaging Assembly (AIA) on board
the forthcoming Solar Dynamic Observatory.

According to these considerations, a hard detection of hot plasma may
point to a scenario of coronal loops heated by single short and relatively
intense nanoflares. Very short durations and very long repetition times
naturally imply that the pulses should be deposited always in different
strands and that, therefore, the strands involved should be a very large
number, i.e. the loop must have a very fine transversal structure.
A scenario of loops heated by short nanoflares has some important
implications.  Hot plasma may be difficult to detect not because of
a limitation of the telescopes, detectors and filters, but due to an
intrinsic property of the plasma emission.  The critical parameter
for detectability of hot plasma is the duration of the heat pulse and
$\sim 1$ minute is the critical duration value, a useful indication for
detailed models of nanoflaring mechanisms. The fine temporal structure
would also automatically imply a very fine spatial structure, difficult
to resolve.


This picture is fully compatible with the current scenario
of nanoflaring multi-stranded coronal loops. For instance, it
naturally involves the hot-underdense/cool-overdense loop cycle
(\citealt{2002ApJ...579L..41W},\citealt{2004ApJ...605..911C},
\citealt{2006SoPh..234...41K}). Less trivial is the effect on the
diagnostics of the detailed thermal structure along and across the loops
(e.g., \citealt{2004ApJ...605..911C}, \citealt{2005ApJ...633..499A},
\citealt{2006A&A...449.1177R}, \citealt{2007ApJ...658L.119S}), which we
defer to later work.

In conclusion, we have investigated in detail the effect of the
non-equilibrium of ionization driven by nanoflares in finely structured
coronal loops on the detection of hot plasma in the loops, and provided
constraints on the heat pulse duration which may make or make not the
detection difficult. Further more detailed information
and diagnostics are expected from high resolution spectroscopy such
as that from EUV Imaging Spectrometer on board the Hinode mission or
from the Extreme Ultraviolet Variablity Experiment (EVE) on board the
forthcoming Solar Dynamic Observatory.

\bigskip
\acknowledgements{We thank G. Peres and the referee, J. Klimchuk,
for suggestions. 
The software used in this work was
in part developed by the DOE-supported ASC/Alliance Center for
Astrophysical Thermonuclear Flashes at the University of Chicago,
using modules for non-equilibrium ionization, thermal conduction,
and optically thin radiation built at the Osservatorio Astronomico
di Palermo. The simulations were executed at the SCAN facility of the
Osservatorio Astronomico di Palermo.  This work was supported by
Ministero
dell'Istruzione, dell'Universit\`a e della Ricerca, by Istituto
Nazionale
di Astrofisica, and by Agenzia Spaziale Italiana
(ASI), contract I/015/07/0. FR acknowledges support
from the International Space Science Institute in the framework of
an international working team.}

\appendix

\section{Non-equilibrium ionization in the FLASH code}

The non-equilibrium ionization is added to the FLASH code using a
method of time splitting between the hydrodynamic and the NEI numerical
modules. A fractional step method is required to integrate the equations
and in particular to decouple the NEI solver from the hydrodynamic solver.
For each timestep, the homogeneous hydrodynamic transport equations given
by Eqs. \ref{eq1}--\ref{eq4} are solved using the FLASH hydrodynamic
solver with R = 0. After each transport step, the  stiff system of
ordinary differential equations for the NEI problem with the form:

\begin{equation}
\frac{\partial n_i^Z}{\partial t} = R_i^Z ~~~~~~~~ (i = 1, ...,
N_{spec})
\end{equation}

\noindent
are integrated. This step incorporates the reactive source terms.
Within each grid cell, the above equations can be solved separately with
a standard ODE method. Since this system is  stiff, it is solved using
the Bader-Deuflhard time integration solver\footnote{The variable-order
Bader-Deuflhard routine used here is a combination of the routine METANI
given by \cite{bader} and the routine STIFBS given by \cite{press}
(see also \citealt{1999ApJS..124..241T}).}  with the MA28 sparse matrix
package. \cite{1999ApJS..124..241T} has shown that these two algorithms
together provide the best balance of accuracy and overall efficiency.
Note that the source term in the NEI module is adequate to solve the
problem for optically thin plasma in the coronal  approximation; just
collisional ionization, auto-ionization, radiative recombination, and
dielectronic recombination are considered. 

\bibliographystyle{apj}
\bibliography{references}

\clearpage


\clearpage

\clearpage


\end{document}